\documentstyle[12pt]{article}
\newcommand{\be}{\begin{equation}}
\newcommand{\ee}{\end{equation}}
\newcommand{\bea}{\begin{eqnarray}}
\newcommand{\eea}{\end{eqnarray}}
\newcommand{\ba}{\begin{array}}
\newcommand{\ea}{\end{array}}

\def\bbox{{\,\lower0.9pt\vbox{\hrule \hbox{\vrule height 0.2 cm
\hskip 0.2 cm \vrule height 0.2 cm}\hrule}\,}}
\newcommand{\dsl}{\pa \kern-0.5em /}

\newcommand{\nn}{\nonumber \\}

\font\mybb=msbm10 at 10pt
\def\bb#1{\hbox{\mybb#1}}
\def\bZ {\bb{Z}}
\def\bR {\bb{R}}
\def\bE {\bb{E}}

\def\bC {\bb{C}}

\input epsf.tex

\begin{document}


\begin{titlepage}
\vfill
\begin{flushright}
DAMTP-1999-165\\
hep-th/9911154\\
\end{flushright}

\vfill

\begin{center}
\baselineskip=16pt
{\Large\bf PhreMology: calibrating M-branes\footnote{To appear in the
proceedings of `Strings 99'.}}
\vskip 0.3cm
{\large {\sl }}
\vskip 10.mm
{\bf Paul K. Townsend} \\
\vskip 1cm
{\small
DAMTP, University of Cambridge, \\
Silver Street, Cambridge CB3 9EW, UK
}
\end{center}
\vfill
\par
\begin{center}
{\bf ABSTRACT}
\end{center}
\begin{quote}
The relevance of calibrations, and `generalized' calibrations, to
supersymmetric M-brane configurations, and their associated field theories, is
reviewed, with emphasis on applications to domain walls and domain
wall junctions of D=4 N=1 supersymmetric field theories.

\vfill

\end{quote}
\end{titlepage}
\setcounter{equation}{0}

Minimal surfaces have long been an active area of mathematics. With the advent
of the `brane revolution' they have become important to theoretical particle
physics via M-theory. An important development on the mathematical side was the
introduction in 1982 of the concept of a {\sl calibration} \cite{HL}. In the
simplest cases this deals with $p$-dimensional submanifolds of $\bE^n$. A
p-form $\phi$ on $\bE^n$ is a calibration if, for all tangent p-planes $\xi$, 

\noindent
(i)  $\phi_\xi \le vol_\xi $

\noindent
(ii) $d\phi =0$

\noindent 
The contact set of the calibration is the set of all tangent p-planes for which
the calibration inequality (i) is saturated, and a calibrated p-surface is one
for which all tangent p-planes belong to the contact set of a
calibration\footnote{Each p-plane tangent to a surface in $\bE^n$ 
corresponds to a point on the  Grassmannian of p-planes through the origin of
$\bE^n$, so the contact set of a calibration is a subset of this Grassmannian.
This relies on the fact that $\bE^n$ has trivial holonomy, but a modified theory
exists for spaces of reduced holonomy.}. A theorem of Harvey and  Lawson states
that a calibrated p-surface has minimum p-volume among all p-surfaces with the
same homology
\cite{HL}. That is, it is a minimal p-surface. 

The proof of this theorem is elementary. Consider an open set $U$ of a surface
satisfying the premise. Since the calibration
inequality (i) is {\sl saturated}, by hypothesis, we have
\be
{\rm vol}(U) = \int_U\! \phi 
\ee
Now consider any other p-surface that coincides with the original one on the
boundary $\partial U$ of $U$, and let $U'$ be an open set of this new 
surface with $\partial U' =\partial U$. Since we only consider surfaces within
the same homology class there exists a $(p+1)$-surface $D$ such that $\partial
D=U-U'$. It follows that
\be
{\rm vol}(U) = \int_{U'}\! \phi + \int_D d\phi\, .
\ee
By property (i) the first term on the right hand side cannot exceed
${\rm vol}(U')$, while the second term vanishes by property (ii).
We therefore deduce the inequality
\be
{\rm vol}(U) \, \le \, {\rm vol}(U')\, .
\ee
The theorem then follows from the fact that this is true 
for any choice of $U$.
 
Applications of this theorem to branes, pioneered\footnote{No attempt will be
made here to survey the many subsequent applications, except those of direct
relevance to D=4 N=1 domain walls and their intersections to be discussed
below.} in \cite{BBMOOY,SS,GPap,GLW,AFS},  
rely on the fact that, in many cases of interest, the energy of a static p-brane
is proportional to the p-volume of its worldspace
$w$, the constant of proportionality being the p-volume tension $T$. There are,
however, many cases in which the energy is not simply the p-volume. Obvious
examples are D-branes and the M5-brane which have gauge-fields on their
worldvolume. The theory of calibrations is still applicable in these cases if
(when consistency allows it) one restricts attention to configurations for which
worldvolume gauge fields vanish. Discounting worldvolume gauge fields, the
bosonic p-brane action takes the universal form
\be
S= -T\int_W \left[vol(g) + A\right]
\ee
where $vol(g)$ is the (p+1)-volume form in the metric $g$ induced on the
worldvolume $W$ swept out by the worldspace $w$ in the course of its time
evolution, and $A$ is a (p+1)-form induced on $W$ from a background (p+1)-form.
For stationary backgrounds one has a timelike Killing 
vector field $k$ for which ${\cal L}_kF$ vanishes, where
$F=dA$ is the (p+2)-form field strength for $A$.  
This Kiling vector field generates a symmetry of
the p-brane action for which there is a corresponding Noether energy. For
static branes\footnote{See \cite{AFSS} for a discussion of calibrations in
relation to non-static branes.} in static backgrounds the energy density is
\be\label{hamden}
{\cal H} = T\left[\sqrt{-k^2}\sqrt{\det m} + \Phi\right]
\ee
where $m$ is the induced worldspace metric, and $\Phi$ is the
worldspace dual of $i_kA$ (in a gauge for which ${\cal L}_kA=0$). In
other words, $\Phi$ is an `electrostatic' energy density 
associated with the background gauge potential $A$. 
If $\Phi$ vanishes
then a minimal energy p-brane is a minimal p-surface (in an appropriately
rescaled metric unless $k^2\equiv -1$). If $\Phi$ does not vanish then a brane
of minimal energy is not a minimal surface. However, in 
this case one can invoke a theory of 
`generalized calibrations' \cite{GutP,GPT} 
(not to be confused with generalizations involving non-vanishing {\sl
worldvolume} gauge fields \cite{GLW2,JPG}). 

A `generalized calibration' is a p-form satisfying, for all tangent p-planes
$\xi$, 

\noindent
(i)' $\phi_\xi \le \sqrt{-k^2}\, vol_\xi$

\noindent
(ii)' $d\phi = i_kF$

\noindent
For simplicity we shall assume here that $k^2\equiv -1$, and refer to
\cite{GPT} for the general case. In this case, property (i)' reduces to
property (i) and the same arguments as before lead to
\be
{\rm vol}(U) \ \le \ {\rm vol}(U') + \int_D d\phi\, .
\ee
In the gauge for which ${\cal L}_kA=0$ we have $i_kF= -d(i_kA)$ and
hence, from property (ii)', $d\phi = -d(i_kA)$. Thus,  
\be
\int_D d\phi = -\int_U\! i_kA + \int_{U'}\! i_kA\, ,
\ee
and we deduce the generalized bound
\be\label{genmin}
{\rm vol}(U) + \int_Ui_kA \ \le \ {\rm vol}(U') + \int_{U'}i_kA\, .
\ee
This is equivalent to
\be
E(U) \le E(U')
\ee
where $E$ is the integral of the energy density ${\cal H}$ of (\ref{hamden}),
with $k^2\equiv -1$. This illustrates the general result of \cite{GPT} that the
contact set of a generalized calibration is a minimal {\sl energy} p-surface.

Generalized calibrations are needed to study supersymmetric branes in 
those supergravity backgrounds for which the `electrostatic' energy density
$\Phi$ is non-vanishing. In this contribution I will limit myself to 
supergravity backgrounds for which $\Phi$ vanishes, and for 
which the background metric is flat, so that $k^2\equiv-1$. In such
cases only the
standard calibrations are needed. I will begin by explaining how, in these
simple circumstances, the calibration bound (i) follows 
from the p-brane supersymmetry algebra; I refer
the reader to \cite{GPT} for the general case. A super p-brane in a vacuum
background is invariant under supertranslations of superspace. This invariance
implies the existence of spinorial Noether charges $Q$, in addition to the
energy and momentum. As this symmetry is a rigid one, the charges in any 
region $U$ of the brane are well-defined. When account is taken of the p-form
central charge in the spacetime supertranslation algebra \cite{az}, one finds
that
\be\label{alg}
\{Q,Q\} = \int_U \left[vol\, \pm \Gamma_0 \Gamma_{I_1\dots I_p}
dX^{I_1}\wedge\dots\wedge dX^{I_p}\right]
\ee
where $vol$ is the volume p-form in the induced {\sl worldspace} 
metric $m$, and $X^I$ are the n-space coordinates. The sign 
depends on the orientation of the p-brane in $\bE^n$. 
The values of $p$ and $n$ are restricted by
supersymmetry, but these restrictions are those required anyway for
applications to M-theory. The values of $p$ and $n$ 
are further restricted if we
assume that the supercharges $Q$ are real. This assumption is not essential  
but will be made here to simplify the presentation. The matrices
$(\Gamma_0,\Gamma_I)$ are the Dirac matrices of (1+n)-dimensional Minkowski
spacetime. 

We now introduce a real, commuting, {\sl covariantly constant}
spinor $\epsilon$, to be
called a `Killing spinor', normalized so that
\be\label{norm}
\epsilon^T\epsilon =1\, .
\ee
The number of such spinors will always equal the number of 
supersymmetry charges $Q$. Given 
such a spinor, (\ref{alg}) implies that
\be\label{pos}
\left(Q\epsilon\right)^2 = \int_U \left[vol  \pm \phi\right]\, ,
\ee
where
\be\label{cali}
\phi = {1\over p!}\left(\epsilon^T\Gamma_0
\Gamma_{I_1\cdots I_p}\epsilon\right)\, 
dX^{I_1}\wedge \cdots \wedge dX^{I_p}\, .
\ee
The left hand side of (\ref{pos}) is manifestly positive, and since this is
so for any region $U$ we must have
\be\label{calbound}
\phi_\xi \le vol_\xi
\ee
for all $\xi$. It is also obvious, since $\epsilon$ is covariantly constant,
that $d\phi=0$. We conclude that $\phi$ is a p-form calibration. 
As we shall now
see, the calibration inequality (\ref{calbound}) is saturated 
by configurations that preserve some fraction of the spacetime supersymmetry. 

The calibration just found from considerations of supersymmetry is
such that
\be
\phi_\xi = vol_\xi\, (\epsilon^T \Gamma_\xi \epsilon)
\ee
where $\Gamma_\xi$ is the matrix
\be
\Gamma = {1\over p! \sqrt{\det m}}\, \varepsilon^{i_1\dots i_p}
\left(\partial_{i_1}X^{I_1}\dots \partial_{i_p}X^{I_p}\right)\,
\Gamma_0\Gamma_{I_1\dots I_p} \, ,
\ee
evaluated at the point to which the p-plane $\xi$ is tangent. 
Given the restrictions on $p$ and $n$ mentioned previously, it can be
shown that
\be
\Gamma^2=1\, .
\ee
The eigenvalues of $\Gamma$ are therefore $\pm 1$, and the 
bound (\ref{calbound}) is an
immediate consequence of this. As we have already derived this bound, the more
relevant point here is that the condition for its saturation is 
\be\label{susycon}
\Gamma\epsilon=\epsilon\, .
\ee 
This is the key equation for what follows\footnote{The Lagrangian
version of this equation was originally 
derived in \cite{bergs,BBS} from considerations of `kappa-symmetry'. The
derivation here follows \cite{GPT}.}. For many simple 
applications we may choose
coordinates for which the Killing spinor $\epsilon$ is constant. 
For simplicity, let us suppose that such coordinates have been chosen and that
$\epsilon$ is constant. For a given tangent p-plane $\xi$, equation 
(\ref{susycon}) becomes $\Gamma_\xi \epsilon=\epsilon$, which states
that $\epsilon$ must belong to the $+1$ eigenspace of
$\Gamma_\xi$; let us call this $S^+_\xi$. This space has a
dimension equal to half the number of supersymmetry charges so, {\sl
locally}, the brane preserves 1/2 supersymmetry. This will also be true
globally if the brane geometry is planar, but in general $\xi$ will
depend on position on the brane and hence $S^+_\xi$ will vary with
position. In this case  the space of solutions
of (\ref{susycon}) is the intersection of 
the spaces $S^+_\xi$ for all $\xi$. For a generic p-brane this space
is the empty set but for special cases there will be a non-empty intersection. 
Non-planar branes (or intersections of planar branes) for which
(\ref{susycon}) has at least one solution are calibrated by $\phi$, but 
will generally preserve less than 1/2 supersymmetry. As we go from one
point on the p-surface to another we go from one p-plane $\xi$ to another
p-plane $\xi'$. Correspondingly,
\be
\Gamma_\xi \rightarrow \Gamma_{\xi'} = R^{-1}\Gamma_\xi R\, ,
\ee
where $R$ is some $SO(n)$ rotation matrix in the spinor representation. The
p-surface will be a calibrated one only if there exist non-zero solutions
$\epsilon$ to $R\epsilon=\epsilon$, because only in this case will $S^+_\xi$ 
and $S^+_{\xi'}$ have a non-empty intersection. 

In the context of M-theory, $\epsilon$ is a 32 component Majorana spinor of
$SO(1,10)$, which is real in a real representation of the Dirac matrices. For
static solutions we may consider $\epsilon$ to be 
a spinor of $SO(10)$, so $n\le 10$.
Preservation of some non-zero fraction $\nu$ of supersymmetry by a static
M-brane configuration requires $R$ to take values in a subgroup $G$ of
$SO(n)\subset SO(10)$ such that the decomposition of the spinor representations
of $SO(n)$ contained in the spinor of $SO(10)$ includes 
at least one singlet. If
the total number of singlets is $32\nu$ then the 
configuration will preserve the
fraction $\nu$ of supersymmetry;  it 
is `$\nu$-supersymmetric'. The subgroups of
$SO(10)$ with the required property are 
\bea
SU(5)   &\subset SO(10) \nonumber \\
Spin(7) &\subset SO(8) \nonumber \\
SU(4)   &\subset SO(8) \nonumber \\
G_2     &\subset SO(7) \nonumber \\
SU(3)   &\subset SO(6) \nonumber \\
SU(2)   &\subset SO(4) 
\eea
For each supersymmetric, and hence calibrated, p-surface the tangent planes
must parameterize a coset space $G/H$, where $G$ is the rotation group
discussed above and $H$ is some stability subgroup. The groups $G$ and $H$,
together with the dimension $p$ of the calibrated surface, provide a
classification \cite{GPap,GLW,AFS} of the calibrations relevant to 
supersymmetric configurations of M5-branes\footnote{There are additional
cases when one includes M2-branes \cite{GPap,GPT}. See \cite{JFF} for a
comprehensive review.}. These are shown in Table 1. 
\begin{table}[hbt]
\caption{Calibrations and supersymmetry}
\small
\begin{tabular}{|c|c|c|c|c|c|}
\hline
p & n & $G$ & $H$ & Calibration Type & Susy fraction \\
\hline
2 & 4 & $SU(2)$ & $U(1)$ & K{\"a}hler & $1/4$\\
\hline
2 & 6 & $SU(3)$ & $S[U(2)\times U(1)]$ & K{\"a}hler & $1/8$\\
\hline
4 & 6 & $SU(3)$  & $S[U(2)\times U(1)]$ & K{\"a}hler & $1/8$\\
\hline
3 & 3 & $SU(3)$ & $SO(3)$ & Special Lagrangian & $1/8$ \\
\hline
3 & 7 & $G_2$   & $SO(4)$ & Associative & $1/16$\\
\hline
4 & 7 & $G_2$ & $SO(4)$ & Co-associative & $1/16$\\
\hline
4 & 8 & $SU(4)$ & $S[U(2)\times U(2)]$ & K{\"a}hler & $1/16$\\
\hline
4 & 8 & $SU(4)$ & $SO(4)$ & Special Lagrangian & $1/16$ \\
\hline
4 & 8 & $Spin(7)$ & $[SU(2)\times SU(2)\times SU(2)]/\bZ_2$ & 
Cayley & $1/32$ \\
\hline
5 & 10 & $SU(5)$ & $SO(5)$ & Special Lagrangian & $1/32$ \\
\hline
\end{tabular}
\end{table}

The simplest realization of these possibilities is 
via orthogonally intersecting
M5-branes. Here I will follow the approach 
of \cite{GLW}. Consider, for example,
two M5-branes intersecting according to the array
$$
\begin{array}{lccccccccccc}
M5: & 1 & 2 & 3 & | & 4 & 5 & - & - & - & - & - \nn
M5: & 1 & 2 & 3 & | & - & - & 6 & 7 & - & - & - 
\end{array}
$$
Each M5-brane determines a tangent p-plane $\xi$, 
and hence an associated matrix
$\Gamma_\xi$ with $+1$ eigenspace $S^+_\xi$. The intersection of these spaces
can be shown, by standard methods, to be 8-dimensional, so the configuration
preserves 1/4 supersymmetry and must be a calibrated configuration. 
The 1-2-3 directions, separated from the others in the above array by the
vertical line, are common to both M5-brane worldvolumes. They play an
inessential role and may be ignored, as may the transverse 8-9-$\natural$
directions (following \cite{nobu}, we use `$\natural$' as a convenient single
character for `ten'), so we effectively have two intersecting 2-branes in
$\bE^4$.  The 2-brane in the 6-7 transverse directions may be
considered as a `solitonic' deformation of a test brane in the 4-5 directions.
From this perspective, the orthogonal intersection of the two 2-branes is a
singular limit of a configuration of a single 2-brane, which can be considered
as an elliptic curve in $\bC^2$, asymptotic to the 
two orthogonal 2-planes in the
4-5 and 6-7 directions. Either plane can be obtained from the other
by a discrete $SU(2)$ rotation in $\bC^2$. Such rotations preserve 1/4
supersymmetry \cite{BDL}. We can desingularize the intersection, while
maintaining 1/4 supersymmetry by allowing a 
continuous $SU(2)$ rotation from one
asymptotic 2-plane to another. We then have a non-singular elliptic
curve which, because it preserves 1/4 supersymmetry, must be a calibrated
2-surface. It must be calibrated by an $SU(2)$ K{\"a}hler 
calibration as this is
the only case with the required properties, the K{\"a}hler and Special
Lagrangian calibrations being equivalent for $SU(2)$. 

This conclusion can be confirmed directly from the constraints imposed
on Killing spinors by two orthogonally intersecting
M5-branes \cite{GLW}. Let us consider here the simpler case of two
orthogonally intersecting M2-branes because the three
common directions of the two M5-branes are irrelevant 
to the final result. We take the M2-branes to intersect according to
the array
$$
\begin{array}{lcccccccccc}
M2: & 1 & 2 & - & - & - & - & - & - & - & - \nn
M2: & - & - & 3 & 4 & - & - & - & - & - & - 
\end{array}
$$
The calibration form (\ref{cali}) in this case is the 2-form
\be
\phi =  {1\over2}\sum_{i,j=1}^4 \left(\epsilon^T\Gamma_{0ij}\epsilon
\right)\, dx^i\wedge dx^j \, .
\ee
The constraints imposed on the Killing spinors by the two M2-branes are 
\be\label{constraints}
\Gamma_{012}\epsilon =\epsilon\, ,\qquad 
\Gamma_{034}\epsilon =\epsilon\, .
\ee
These imply 
\be
\Gamma_{12}\epsilon = \Gamma_{34}\epsilon \, ,\qquad 
\Gamma_{13}\epsilon = -\Gamma_{24}\epsilon \, ,\qquad
\Gamma_{14}\epsilon = \Gamma_{25}\epsilon\, ,
\ee
and also
\be
\epsilon^T\Gamma_{013}\epsilon =0\, \qquad 
\epsilon^T\Gamma_{014}\epsilon =0\, .
\ee
Given the normalization (\ref{norm}) of $\epsilon$, we then find that
\be
\phi = dX^1\wedge dX^2 + dX^3\wedge dX^4\, .
\ee
Introducing the complex coordinates $z=X^1+iX^2$ and $w=X^1+iX^2$ we see that 
$\phi$ is a closed hermitian 2-form on $\bE^4$; that is, a K{\"a}hler 2-form.
It is also $SU(2)$-invariant with $(dz,dw)$ transforming as a complex doublet. 
The 2-form $\phi$ is therefore an $SU(2)$ K{\"a}hler calibration.

The same conclusion can be reached for M5-branes via a similar
analysis of the constraints imposed on Killing spinors (after
factoring out from the 5-form of (\ref{cali}) the volume 3-form of the
three common directions). But these constraints apply to a much wider
class of configurations than orthogonally intersecting M5-branes.
In particular, they apply to any 1/4 supersymmetric configuration of a
single M5-brane that is asymptotic to the configuration of orthogonally
intersecting M5-branes. The asymptotic planes may also be rotated relative to
each other (see \cite{AFS} for a discussion of calibrations in this context). 
One may consider the array as a shorthand for the entire collection of
such configurations. Pushing this interpretation to the extreme, one may regard
the array as {\sl nothing more than a tabular representation of the
constraints} (\ref{constraints}), so that it may be 
considered to represent any
configuration yielding these constraints regardless of the asymptotic
behaviour. If an interpretation this liberal 
is adopted then the geometry of the 
array may suggest results that are spurious in particular contexts, but one may
still hope to capture general features common to all cases. We shall
return to this point below when discussing the associative and Cayley
calibrations, but we first need to consider a simpler case with 1/8
supersymmetry. 

Consider three M5-branes intersecting according to the array 
$$
\label{array2}
\begin{array}{lccccccccccc}
M5: & 1 & 2 & 3 & | & 4 & 5 & - & - & - & - & - \nn
M5: & 1 & 2 & 3 & | & - & - & 6 & 7 & - & - & - \nn
M5: & 1 & 2 & 3 & | & - & - & - & - & 8 & 9 & - 
\end{array}
$$
Each M5-brane determines a tangent p-plane $\xi$, and hence an associated matrix
$\Gamma_\xi$ with $+1$ eigenspace $S^+_\xi$. The intersection of these spaces
can be shown, by standard methods, to be 4-dimensional, so the configuration
preserves 1/8 supersymmetry and must be a calibrated one. The 1-2-3
directions, separated from the others in the above array by the vertical line,
are common to all M5-brane worldvolumes. They play an inessential role and may
be ignored, as may the transverse tenth direction, so we effectively have three
intersecting 2-branes in $\bE^6$. The 2-branes in the 6-7 and 8-9 transverse
directions may be considered, as before, as `solitonic' deformations of a test
brane in the 4-5 directions. In this case the orthogonal intersection of the
three 2-branes can be considered a singular limit of a configuration of a single
2-brane `wrapped' on an elliptic curve in $\bC^3$, and asymptotic to the three
orthogonal 2-planes in the 4-5, 6-7, and 8-9 directions. Each of these planes can
be obtained from any of the other two by a discrete $SU(3)$ rotation in
$\bC^3$. Such rotations preserve 1/8 supersymmetry \cite{BDL}. We can
desingularize the intersection, while maintaining 1/8 supersymmetry by allowing
a continuous $SU(3)$ rotation from one asymptotic 2-plane to another. We then
have a non-singular elliptic curve which, because it preserves 1/8
supersymmetry, must be a calibrated 2-surface. It must be calibrated by
an $SU(3)$ K{\"a}hler calibration as this is the only candidate with the 
required properties. This can be verified as before from the constraints
imposed on Killing spinors by the intersecting M5-brane configuration. 

In any configuration of intersecting branes there will generally be zero modes
trapped on the intersection and these will govern its low energy
dynamics. In the above example the intersection is 3-dimensional
so the zero modes trapped on the intersection yield a
(3+1)-dimensional quantum field theory. Since the 
configuration preserves 1/8 of the supersymmetry of the M-theory
vacuum, the intersection field theory has a total of four
supersymmetries, transforming as a real spinor of $SO(1,3)$. In other
words, the low energy intersection dynamics is governed by a D=4 N=1
SQFT. If the intersection is desingularized so as to describe a single
non-singular M5-brane then this SQFT will be determined by 
the M5-brane's effective action. The SQFT obtained in this way
from three orthogonally-intersecting
M5-branes is not of particular interest in itself, but any minimal energy
M5-brane wrapping a Riemann surface in a flat M-theory vacuum 
is also calibrated
by an $SU(3)$ K{\"a}hler calibration, and various SQFTs can be thus
obtained.
A particularly interesting example in which an M5-brane wraps a Riemann
surface in the $S^1$-compactified M-theory vacuum was identified in \cite{wit}
as a theory, now called `MQCD', with properties similar to that of SQCD. 

An example of an associative calibration is
provided by four M5-branes intersecting according to the array \cite{GLW}
$$
\begin{array}{lccccccccccc}
M5: & 1 & 2 & | & 3 & 4 & 5 & - & - & - & - & - \nn
M5: & 1 & 2 & | & 3 & - & - & 6 & 7 & - & - & - \nn
M5: & 1 & 2 & | & 3 & - & - & - & - & 8 & 9 & - \nn
M5: & 1 & 2 & | & - & 4 & - & 6 & - & 8 &  & - 
\end{array}
$$ 
Ignoring the common directions, and the transverse tenth direction, and
interpreting the last three rows as `solitonic' deformations of a test 3-brane
in the 3-4-5 directions, we conclude that we have a calibrated 3-surface in
$\bE^7$. Since this configuration preserves 1/16 supersymmetry, it must be
calibrated by a 3-form, and the associative 3-form calibration is the only
candidate. We might wish to relate this to a D=3 SQFT, but there is an
alternative application. Noting that this array can be obtained from
the previous one by the addition of a fourth M5-brane, we retabulate it as
$$
\begin{array}{lccccccccccc}
M5: & 1 & 2 & 3 & | & 4 & 5 & - & - & - & - & - \nn
M5: & 1 & 2 & 3 & | & - & - & 6 & 7 & - & - & - \nn
M5: & 1 & 2 & 3 & | & - & - & - & - & 8 & 9 & - \nn
    & - & - & - & | & - & - & - & - & - & - & - \nn
M5: & 1 & 2 & - & | & 4 & - & 6 & - & 8 & - & -  
\end{array}
$$
If the first three M5-branes are interpreted as supplying a D=4 N=1 SQFT with
coordinates $(x^0,x^1,x^2,x^3)$ then the array suggests that we interpret the
fourth M5-brane as a domain wall in this theory. The 1/16 supersymmetry of
associative calibrations of an M5-brane translates to 1/2 supersymmetry of
the D=4 N=1 Minkowski vacuum; in other words, the domain wall is a 1/2
supersymmetric `BPS' wall. This is a simple analogue of Witten's
identification of associative calibrations of the M5-brane of MQCD as 1/2
supersymmetric domain walls \cite{wit} (see also \cite{KSY,volovich}). It
has been known for some time \cite{AT,cvetic} that the WZ model admits 1/2
supersymmetric domain walls for an appropriate superpotential (the solutions
having been studied originally as solitons of the dimensionally-reduced N=2 D=2
SQFT \cite{FMVW}).  They also appear in $SU(n)$ SQCD with $n\ge2$ \cite{DS}
because the low effective action includes a WZ action with a superpotential
admitting $n$ isolated critical points and hence $n$ degenerate vacua \cite{VY}.
Minimal energy configurations that interpolate from one vacuum to another are 1/2
supersymmetric domain walls. It should therefore be no surprise that MQCD 
admits similar domain walls, although their geometrical nature (and the fact
that they are D-branes \cite{wit}) is remarkable. 

It has recently been appreciated that the domain walls of WZ models, and
hence of SQCD, may intersect at junctions, the configuration as a whole
preserving 1/4 supersymmetry \cite{GT,CHT}. The junction of two domain walls is
1-dimensional. Let $z$ be a complex coordinate for the plane orthogonal to this
direction. The 1/4 supersymmetric domain wall junctions of the WZ model are
configurations $\psi(z)$ of a complex scalar field satisfying
\be\label{junction}
{d\psi\over d z} = \overline{W'(\psi)}\, ,
\ee
where the `superpotential' $W$ is a holomorphic function of $\psi$. 
The 1/2 supersymmetric domain walls themselves are special solutions of this
equation with translational symmetry along one direction in the $z$-plane, 
but the generic solution is 1/4 supersymmetric. In the case that 
\be\label{superp}
W' = 1 - \psi^3\, ,
\ee
corresponding to a quartic superpotential, there are three vacua with
$\psi=(1,\omega,\omega^2)$ where $\omega$ is a cube-root of unity.
In this case we expect a $\bZ_3$-symmetric junction, as shown in
Fig. 1.
 \begin{figure}
\begin{center}
\leavevmode
\epsfxsize=.6\textwidth
\epsfbox{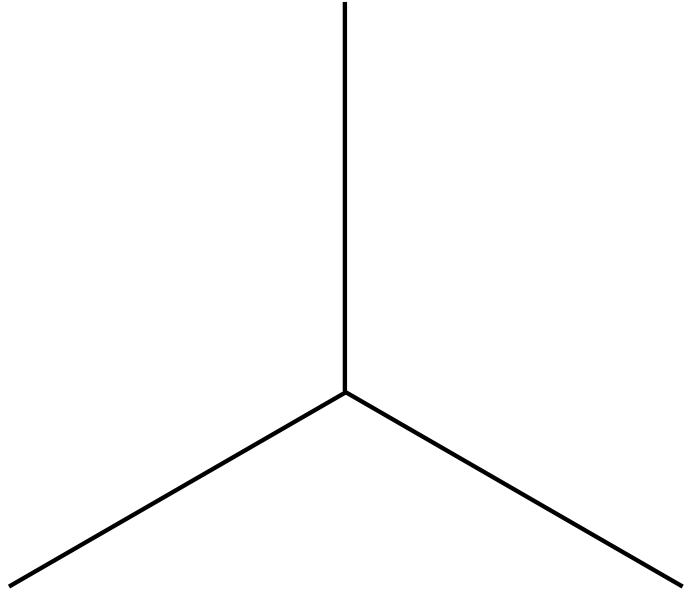}
\bigskip
\caption{Domain wall junction in the complex z-plane}
\label{fig:junk}
\end{center}
\end{figure}
\noindent
Although no exact solution to (\ref{junction}) of this type
is known, numerical studies suggest that one exists
\cite{saffin} and the existence of a $\bZ_3$-invariant minimum energy 
domain wall junction solution to the second order WZ equations 
with superpotential (\ref{superp}) has been proved \cite{math}. In addition, 
an exact solution representing a domain-wall junction has recently
been found in a more complicated, but related, model \cite{Oda}.  
It seems clear from these results that 1/4 supersymmetric 
domain wall junctions are generic to D=4 N=1 SQFTs. In 
particular they are expected in $SU(n)$ 
SQFT for $n\ge 3$ \cite{Gorsky}. 

This raises an obvious question. Does MQCD have 1/4 supersymmetric domain wall
junctions and, if so, what is their geometrical realization? A suggestion
of Gauntlett, which will be explored in a forthcoming article \cite{GGT}, is
that domain wall junctions of MQCD are Cayley calibrations of the MQCD
M5-brane. This can be motivated by considering the realization of Cayley
calibrations as five M5-branes intersecting orthogonally according to the  
array \cite{GLW}
$$
\begin{array}{lccccccccccc}
M5: & 1 & 2 & 3 & | & 4 & 5 & - & - & - & - & - \nn
M5: & 1 & 2 & 3 & | & - & - & 6 & 7 & - & - & - \nn
M5: & 1 & 2 & 3 & | & - & - & - & - & 8 & 9 & - \nn
    & - & - & - & | & - & - & - & - & - & - & - \nn
M5: & 1 & 2 & - & | & 4 & - & 6 & - & 8 & - & - \nn
M5: & 1 & - & 3 & | & 4 & - & 6 & - & - & 9 & -  
\end{array}
$$
This configuration preserves 1/32 supersymmetry, corresponding to 1/4
supersymmetry of the D=4 N=1 vacuum defined by the first three
M5-branes. If the last two M5-branes are viewed as excitations about this
vacuum then it is apparent from the array that they can be interpreted as
intersecting domain walls. 
\vskip 0.5cm

\noindent
{\bf Acknowledgements}: I thank Jerome Gauntlett, Gary Gibbons, Jan 
Gutowski and George Papadopoulos for their collaboration on work
reported here, and for helpful discussions.

\end{document}